\documentclass[aps,prb,reprint,groupedaddress,showpacs]{revtex4-1}

\usepackage{graphicx}% Include figure files
\usepackage{dcolumn}% Align table columns on decimal point
\usepackage{bm}% bold math
\makeatletter
\AtBeginDocument{\@ifpackageloaded{natbib}{\ifNAT@numbers\if@filesw\immediate\write\@auxout{\string\global\string\NAT@numberstrue}\fi\fi}{}}
\makeatother \usepackage{amstext}

\begin{document}

\title{Low-frequency and shot noises in CoFeB/MgO/CoFeB magnetic tunneling
junctions} \author{Tomonori Arakawa$^{1}$} \author{Takahiro Tanaka$^1$, Kensaku
Chida$^1$, Sadashige Matsuo$^1$, Yoshitaka Nishihara$^1$} \author{Daichi
Chiba$^1$} \author{Kensuke
Kobayashi$^{1}$}\email{kensuke@phys.sci.osaka-u.ac.jp} \altaffiliation{Present
address: Graduate School of Science, Osaka University, 1-1 Machikaneyama,
Toyonaka, Osaka 560-0043, Japan.}  \author{Teruo Ono$^1$, Akio Fukushima$^2$}
\author{Shinji Yuasa$^2$}

%\author{***$^{1, \dagger}$, ***$^1$}

\affiliation{$^1$Institute for Chemical Research, Kyoto University, Uji, Kyoto
611-0011, Japan} \affiliation{$^2$Spintronics Research Center, Advanced
Industrial Science and Technology (AIST), AIST Tsukuba Central 2, Tsukuba,
Ibaraki 305-8568, Japan}

\begin{abstract}
The low-frequency and shot noises in spin-valve CoFeB/MgO/CoFeB magnetic
tunneling junctions were studied at low temperature.  The measured $1/f$ noise
around the magnetic hysteresis loops of the free layer indicates that the main
origin of the $1/f$ noise is the magnetic fluctuation, which is discussed in
terms of a fluctuation-dissipation relation.  Random telegraph noise (RTN) is
observed to be symmetrically enhanced in the hysteresis loop with regard to the
two magnetic configurations.  We found that this enhancement is caused by the
fluctuation between two magnetic states in the free layer.  Although the $1/f$
noise is almost independent of the magnetic configuration, the RTN is enhanced
in the antiparallel configuration.  These findings indicate the presence of
spin-dependent activation of RTN.  Shot noise reveals the spin-dependent
coherent tunneling process via a crystalline MgO barrier.

\end{abstract}
\date{\today}% It is always \today, today,
\pacs{75.70.Cn, 73.50.Td, 73.40.Rw,72.25.Ba}

%75.70.Cn Magnetic properties of interfaces (multilayers, superlattices, heterostructures)
%73.50.Td Noise processes and phenomena
%73.40.Rw Metal-insulator-metal structures
%72.25.Ba Spin polarized transport in metals

\maketitle
%\tableofcontents

\section{Introduction}
The magnetic tunneling junction (MTJ), which consists of a tunnel barrier
sandwiched between two ferromagnetic electrodes, is one of the central topics in
spintronics.~\cite{TMR1} MTJs exhibit tunneling magnetoresistance (TMR); their
resistance depends on the relative magnetic configurations (parallel or
antiparallel).  Since the TMR effect was discovered by Julliere,~\cite{TMR2}
amorphous Al$_{2}$O$_{3}$ has been mainly used as a tunnel barrier.~\cite{TMR3,TMR3_add}
However, in 2004, large magnetoresistance (MR) was obtained in MTJs with a
crystalline MgO barrier~\cite{TMR4,TMR5} supported by theoretical
prediction.~\cite{TMR6,TMR7,TMR7_add} These days, MgO-based MTJs are extensively studied
from the viewpoints of fundamental physics and device applications.

Although most MTJ studies have thus far been performed via conventional
resistance measurements, noise measurements can serve to further clarify the
intrinsic properties in MTJs.  
The noise results from the fluctuation of the current (thermal noise and shot noise) and of the resistance such as the $1/f$ noise and the random telegraph noise (RTN).
Thermal noise and shot noise are due to the thermal agitation of electrons and the partition process of electrons, respectively, whereas the resistance fluctuation in MTJs is attributed to a nonmagnetic origin (charge trap in the tunneling barrier) and a magnetic origin (magnetic fluctuations and domain wall motion in the free and/or fixed magnetic layers).
%The noise is the voltage fluctuation between the leads of the sample, resulting from the fluctuation of the current (thermal noise and shot noise) and of the resistance such as the $1/f$ noise and the random telegraph noise (RTN).
%Although current noise is independent of the frequency in the zero-frequency limit, resistance noise is usually obvious in the low-frequency region.

Shot noise offers information on the interactions and/or quantum correlations of
conducting electrons.~\cite{Noise1,Noise2} When the average current $\textit{I}$
is fed to a tunnel junction, the current noise $S_{I}$ resulting from the shot
noise can be expressed as $S_{I}=2eIF$ (in the zero-temperature limit) with Fano
factor $F$.  It is well established that $F=1$ in normal-insulator-normal
junctions,~\cite{Noise3} which means that the electron partition at the junction
obeys a Poissonian process.  In the MTJ case, when the tunnel barrier is
composed of amorphous Al$_{2}$O$_{3}$, electron tunneling can be explained by
the conventional Julliere's model.
However, a coherent tunneling via highly spin-polarized $\Delta _{1}$ Bloch states~\cite{TMR8,TMR9} plays a central role in MTJs with crystalline MgO barriers.
%However, coherent tunneling is theoretically expected to play a central role in MTJs with crystalline MgO barriers.~\cite{TMR8,TMR9}
Although information obtained by conventional $I\text{-}V$ measurements cannot directly address the coherence of electron transport, shot noise offers further insight into the mechanism of electron transport.~\cite{MTJshot1,MTJshot2,MTJshot3,MTJshot4,MTJshot5,MTJshot6,MTJshot8,Othernoise2}
In fact, we reported sub-Poissonian shot noise ($F<1$), which is attributed to
coherent tunneling, in a previous work.~\cite{MTJshot6} Our experimental work is
quantitatively reproduced by the recent theoretical work with first-principles
calculations.~\cite{MTJshot7}

\begin{figure}[bp]%%----------------------------------
\center \includegraphics[width=.90\linewidth]{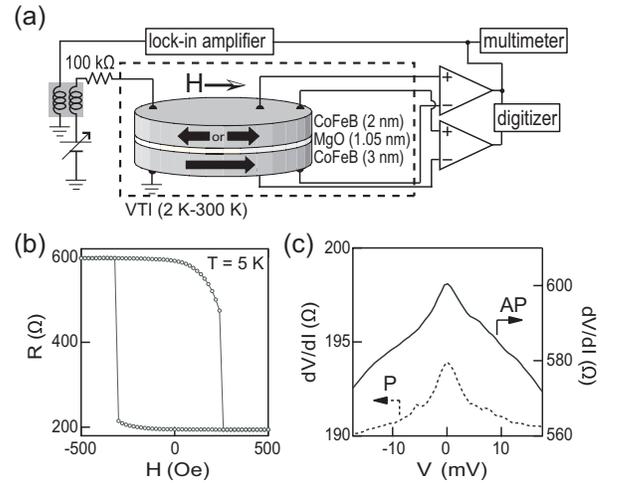} \caption{(a) Schematics
of measurement setup and device structure. (b) Typical MR curves of the present
MTJ measured at 5 K. (c) Typical differential resistance for P (dashed line) and
AP (solid line) configurations as a function of bias voltage.}
\end{figure}

Resistance fluctuations are also important, as MTJs have found broad
application, such as for magnetic field detectors,~\cite{TMRapply1} magnetic
random-access memory,~\cite{TMRapply2,SekiAPL2011} random number generators~\cite{SekiAPL2011}, and microwave
oscillators.~\cite{TMRapply3} For these applications, the signal-to-noise ratio
is critically important, where the $1/f$ noise and RTN limit device performance
at low frequency.
There have been many reports of $1/f$ noise and RTN in MTJs with Al$_{2}$O$_{3}$-based,~\cite{AlOnoise1,AlOnoise2,AlOnoise3,AlOnoise4,AlOnoise5,AlOnoise6,AlOnoise7} MgO-based,~\cite{MgOnoise1,MgOnoise2,MgOnoise3,MgOnoise4,MgOnoise6,MgOnoise6_add,MgOnoise7,MgOnoise8,MgOnoise9,MgOnoise9_add,MgOnoise9_add2,MgOnoise10,MgOnoise11} and other tunneling barriers.~\cite{Othernoise1,Othernoise2}
Nevertheless, little is known about the noise properties of MTJs with submicron-sized junctions with a thin tunneling barrier,~\cite{MgOnoise10,MgOnoise11} which are envisaged for memory and oscillator applications.

In this paper, we report on noise properties, including shot noise, $1/f$ noise,
and RTN in well-crystalline MgO-based MTJs with submicron-sized junctions with
thin tunneling barriers.  The noise measurement was carried out at low
temperature with high experimental accuracy, with a focus on the noise
properties around the magnetic hysteresis loops of the free layer.  We
investigated each noise source systematically as a function of magnetic field and bias
voltage.  The clear dependence of the $1/f$ noise on the applied
magnetic field indicates that the main origin of the $1/f$ noise is magnetic
fluctuation in the free layer, which is discussed in terms of a
fluctuation-dissipation relation.  Based on the bias dependence of RTN, we
discuss the origin of RTN, which is different from that of the $1/f$ noise.  The
analysis of shot noise is also presented to further support our previous
report.~\cite{MTJshot6}

This paper is organized as follows: In Sec.~IIA, we provide information on the
sample fabrication and basic properties of our sample. Then the measurement
system is described.  In Sec.~IIB, the analysis method to extract the
frequency-dependent and frequency-independent components is explained.  Section
III is devoted to the experimental results and discussion.  First, we show the
magnetic field and bias-current dependence of frequency-dependent noise in
Sec.~IIIA, and then the origin of the $1/f$ noise is discussed by using the
fluctuation-dissipation relation in Sec.~IIIB.  We estimate the fluctuating
magnetic moment of the RTN in Sec.~IIIC.  The bias voltage dependence of the
measured RTN is analyzed to obtain information on the excitation mechanism in
Sec.~IIID, and, finally, we show the result of the white noise component in
Sec.~IIIE.  In Sec.~IV, we conclude our study.
\section{Experiment}
\subsection{Device and measurement}
\begin{figure}[tp]%%----------------------------------
\center \includegraphics[width=.82\linewidth]{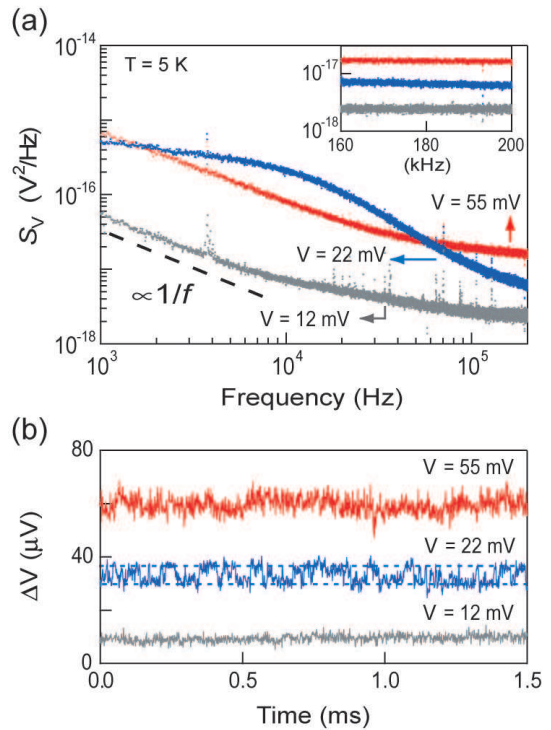} \caption{(a) Typical
voltage noise spectral density of our MTJs in the AP configuration for different
bias voltages: $V=12$, 22, and 55 mV at 5 K. The dashed line is a guide to the
eye. The inset shows the same spectra between 160 and 200 kHz (but note that the
bottom axis is a linear scale). (b) Measured real-time voltage fluctuations
(with vertical offset for clarity) for the same condition as (a). The trace for
$V=22$ mV contains two-level fluctuations, as indicated by the two horizontal
dashed lines.}
\end{figure}
Multilayer stacks of MTJs were deposited in a magnetron sputtering system on a
SiO$_{2}$ layer on a silicon substrate.  The order of the layer structure from
the substrate is as follows; buffer, PtMn(15), CoFe(2.5), Ru(0.85), CoFeB(3),
MgO(1.05), CoFeB(2), and cap, where the top CoFeB layer serves as a free layer
[see Fig.~1(a)].  The thickness of each layer is indicated in ($\ldots$) in
nanometers.
The multilayer stacks are patterned into elliptic pillars with $160 \times 60$ nm dimensions by milling up to the middle of the PtMn layer.  To
crystallize CoFeB layers, the stack is annealed in $10,000$ Oe for 120 min at
330${\,}^\circ$C.~\cite{TMR10,TMR11}

All of the results presented here were obtained at low temperature (3--5 K) in the
variable temperature insert (Oxford VTI) in a magnetic field ($H$) between
$-40,000$ and 500 Oe.  The schematic setup of our measurement system is shown in
Fig. 1(a).  A dc bias is applied under a constant current condition ($I$) by
using the dc voltage source through a 100 k$\Omega$ resistor.  The differential
resistance ($dV/dI$) and the dc bias voltage ($V$) are measured by a lock-in
amplifier and a digital multimeter, respectively.  Figure 1(b) shows the typical
MR curve around the hysteresis loop of the free layer at 5 K.  The clear square
shape of the curve without any steps indicates that there is no pinning site on
a macroscopic domain wall.  The MR ratio defined as $(R_{AP}- R_{P})/R_{P}$ is
208\%, where $R_{P}$ and $R_{AP}$ are the sample resistances in the parallel (P)
and antiparallel (AP) configurations, respectively.  Figure 1(c) shows the
typical bias dependence of the differential resistance, where the solid and
dashed curves correspond to the AP ($-$500 Oe) and P (500 Oe) configurations,
respectively.
In the differential resistance, as temperature decreases below 10 K, a peak structure appears around the zero bias in both P and AP configurations, which is consistent with the several previous reports.~\cite{TMR2,TMR3_add,MTJshot2}
Interestingly, in the P configuration, additional satellite peaks seem to appear around $V$ = $-$5 and 5 mV.
Such a feature may be related to the observation reported before.~\cite{TMRzero}

To obtain the voltage noise spectral density $S_{V}$, we measure the time-domain
voltage fluctuation signal by using a two-channel digitizer (National
Instruments PCI-5922), which yields $S_{V}$ by a fast Fourier transformation.
In this process, two sets of voltage signals are simultaneously measured after
being independently amplified by two room-temperature amplifiers (NF Corporation
LI-75A).  The cross-correlation technique is used here to reduce the external
noise and amplifier noise by long-time averaging.~\cite{CrossCorr} The frequency
range of our system is 100 Hz to 200 kHz. In addition to $S_{V}$, the real-time
voltage signal is also recorded by the digitizer at a sampling rate of 1 MHz.

The $S_{V}$ measurement is carefully calibrated with the thermal noise of
several commercial resistors (MCY100R00T, MCY250R00T, MCY350R00T, and
MCY1K0000T) with a precision of 0.01\%.  The typical resolution of $S_{V}$ for
shot noise estimation is below $ 10^{-20}$~V$^{2}/$Hz.  As a result, we achieved
an experimental precision for the Fano factor well below 1\%.  We measured three
devices with the same geometry (samples 1, 2, and 3) which are made out of the single wafer, and obtained consistent
results.

\subsection{Analysis of noise}
In the present experiment, we found that the voltage noises in MTJs consist of a
white noise ($S_{\rm{white}}$), which is a frequency-independent component,
$1/f$ noise ($S_{1/f}$), and RTN ($S_{\rm{RTN}}$):~\cite{sum_add}
\begin{eqnarray}
S_{V}=S_{\rm{white}}+S_{1/f}+S_{\rm{RTN}}
\label{eq:one}.
\end{eqnarray}
The white noise ($S_{\rm{white}}$) is attributed to the thermal agitation of
electrons (thermal noise) and the partition process of electrons (shot noise).
Then $S_{\rm{white}}$ is described by
\begin{eqnarray}
\nonumber S_{\rm{white}}&=&4k_{B}TR_{d}\\
&+&2F\left[eIR_{d}^{2}\coth \left(\frac{eV}{2k_{B}T}\right)-2k_{B}TR_{d}\right]
\label{eq:two},
\end{eqnarray}
where $k_{B}$ is Boltzmann's constant, $e$ is the electron charge, $R_{d}$ is
the differential resistance ($dV/dI$) at a given $V$ (or $I$), and $F$ is the
Fano factor.  The $1/f$ noise ($S_{1/f}$) in MTJs is parameterized by~\cite{AlOnoise2}
\begin{eqnarray}
S_{1/f}=\frac{\alpha ^{\ast }I^{2}R_{d}^{2}}{{A}f}
\label{eq:three},
\end{eqnarray}
where $\alpha ^{\ast }$, $f$, and $A$ are the Hooge parameter, frequency, and
junction area, respectively.  The RTN is the fluctuation between two levels, where
$S_{\rm{RTN}}$ exhibits a Lorentzian character in an ideal case.

Typical results of $S_{V}$ for the AP configurations for $V=12$, 22, and 55 mV
at 5 K are shown in Fig.~2(a).  In the data, the resistor-capacitor (RC) damping
owing to the capacitance (760 pF) of the measurement lines has already been
corrected for.~\cite{MTJshot6} In the inset of Fig.~2(a), which shows the region
between 160 and 200 kHz, the spectra are almost flat, and thus the increase of
$S_{V}$ with increasing $V$ corresponds to the shot noise.  In contrast, the
spectra for the low-frequency region strongly depend on the frequency, as shown
in the main panel of Fig.~2(a).  The spectra for $V=55$ and 12 mV are clearly
dominated by the $1/f$ noise except for the high-frequency region.  The spectrum
for $V=22$ mV has a clear Lorentzian character,~\cite{MgOnoise11} indicating that the source of
the noise is two-level fluctuation of the resistance, namely, RTN.  Figure~2(b)
shows the real-time voltage signal for $V=12$, 22, and 55 mV.  Apparently, the
time-domain signal has a two-level nature only for $V=22$ mV, as indicated by
the two horizontal dashed lines in Fig.~2(b).

We performed a histogram analysis between 160 and 200 kHz to estimate
$S_{\rm{white}}$.~\cite{MTJshot4} To investigate the $1/f$ noise and RTN, we
first subtract the white noise component from the measured $S_{V}$ and then
define the Hooge parameter $\alpha $ as
\begin{eqnarray}
\alpha ={A}\int_{f_{1}}^{f_{2}}\left(S_{V}-S_{\rm{white}}\right)df/I^{2}R_{d}^{2}\ln \left(\frac{f_{2}}{f_{1}}\right)
\label{eq:four},
\end{eqnarray}
where the frequencies $f_{1}$ and $f_{2}$ are chosen to be 1 and 100 kHz,
respectively.  The value of $\alpha$ thus obtained equals the $\alpha ^{\ast }$
in Eq.~(3) when RTN is negligibly small.  In this study, we mainly use the
$\alpha$ that is a well-defined parameter even if RTN is present.

\section{Results and discussion}
\begin{figure}[tp]%%----------------------------------
\center \includegraphics[width=.94\linewidth]{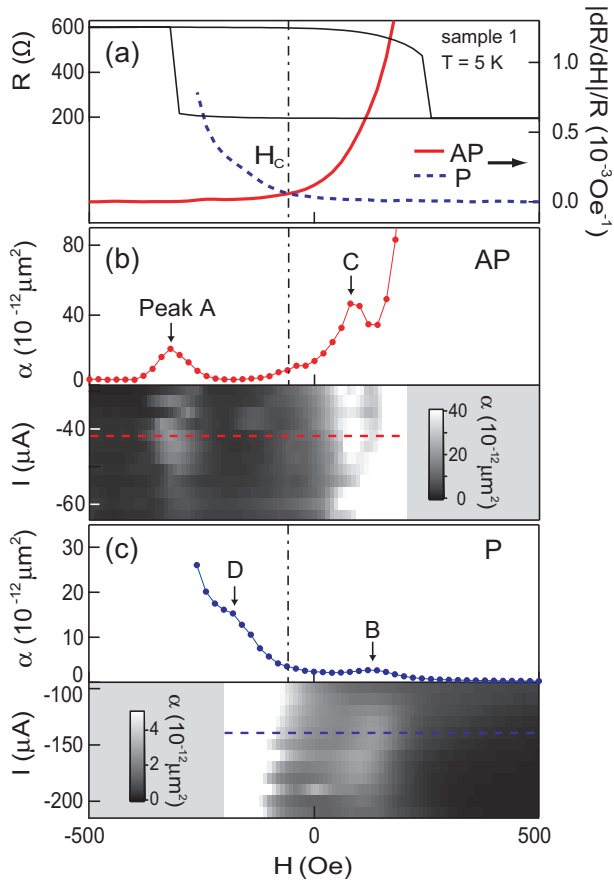} \caption{The magnetic
field and bias current dependence of $\alpha $ measured at 5 K for sample 1. (a)
MR curves of this MTJ and $|dR/dH|/R$ values for P (dashed line) and AP (solid
line) configurations numerically estimated from the MR curves. The vertical
dashed line marks $H=H_{c}$ at which $|dR/dH|/R$ for AP and P have the same
value. (b) The measured $\alpha $ for AP (solid mark) configuration, where we
marked the peaks of $\alpha $ as A and C (upper panel), and a color plot of
$\alpha$ as a function of magnetic field and bias current (lower panel). The
dashed line corresponds to $\alpha $ in the upper panel. (c) Counterpart of
Fig.~3(b) for the P configuration, where the peaks of $\alpha $ are marked as B
and D. $\alpha $ values for the two configurations were measured in almost the
same bias voltage region.}
\end{figure}
\subsection{Frequency-dependent noise}
\begin{figure}[tp]%%----------------------------------
\center \includegraphics[width=.70\linewidth]{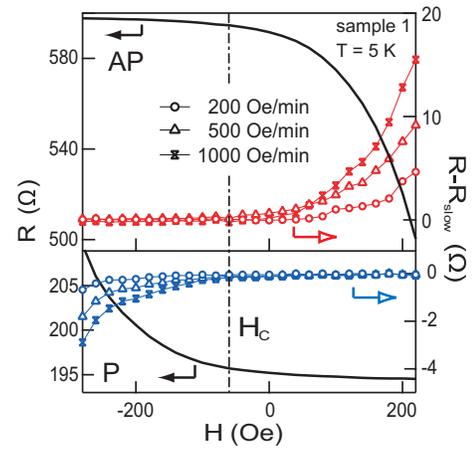} \caption{Field sweep
rate dependence of dc resistance. The solid line is the dc resistance measured
with the slowest sweep rate (50 Oe/min) as the field was ramped from $-500$ to
240 Oe for AP and 500 to $-300$ Oe for P. Results for other sweep rates (200,
500, and 1000 Oe/min) are plotted as differences from the slowest one.}
\end{figure}

We start with the experimental result of the frequency-dependent noise.  There
have been several studies on the field dependence of the $1/f$ noise in
Al$_{2}$O$_{3}$-based~\cite{AlOnoise1,AlOnoise2,AlOnoise3,AlOnoise4,AlOnoise5,AlOnoise7}
and MgO-based~\cite{MgOnoise1,MgOnoise2,MgOnoise3,MgOnoise6,MgOnoise6_add,MgOnoise9,MgOnoise9_add,MgOnoise9_add2,MgOnoise11}
MTJs.  In these reports, the measured $1/f$ noise consists of field-independent
and field-dependent components.  The former component has a nonmagnetic origin
(charge trap in the tunneling barrier), whereas the latter component has a
magnetic origin (magnetic fluctuations and domain wall motion in the free and/or
fixed magnetic layers).  Recent studies on the $1/f$ noise in MgO-based MTJs
have shown that the nonmagnetic $1/f$ noise is no longer important owing to the
improvement of tunneling barrier quality.~\cite{MgOnoise9} For RTN, both
magnetic and nonmagnetic origins were reported in
Al$_{2}$O$_{3}$-based~\cite{AlOnoise1,AlOnoise3,AlOnoise4,AlOnoise5,AlOnoise6} and
MgO-based~\cite{MgOnoise2,MgOnoise4,MgOnoise6,MgOnoise10} MTJs.  Nevertheless,
there are only a few systematic studies on the magnetic field and bias voltage
dependence of RTN.  Here we focus on the magnetic $1/f$ noise and RTN in the
hysteresis loop of the free layer.

Figure~3(a) shows the resistance and $|dR/dH|/R$ at 5 K, which is obtained as
the field is ramped from $-500$ to 200 Oe for AP and from 500 to $-240$ Oe for P
configurations.  The value of $\alpha $ measured at each field for AP and P
configurations is shown in the upper panels of Figs.~3(b) and 3(c),
respectively.
It is clear that the magnetic field dependencies between $\alpha $ and $|dR/dH|/R$ resemble each other for the two configurations:~\cite{MagLoss} They increase
with the reversal of the free layer, except for a few peaks observed in $\alpha$
[peaks A, B, C, and D in Figs.~3(b) and 3(c)].  Whereas the measured spectra are
dominated by the $1/f$ noise outside of these peak regions, strong enhancement
of Lorentzian character is always seen in such peak regions.  The lower panels
of Figs.~3(b) and 3(c) show image plots of $\alpha$ as a function of the
magnetic field and the bias current for the AP and P configurations,
respectively.  In these figures, the ranges of the bias current for the two
configurations are set to be almost the same with respect to the bias voltage.
The curves shown in the upper panel of Figs.~3(b) and 3(c) correspond to the
cross sections indicated by the dashed line in each image plot.

We define $H_{c}$ as the field at which $|dR/dH|/R$ gives the same value for the
AP and P configurations [see Fig.~3(a)].  The behaviors of $\alpha $ and
$|dR/dH|/R$ are found to be well symmetric with respect to $H_{c}$, except for
the peaks.  The values of $\alpha $ for AP and P configurations at their
baselines are $3.3\times 10^{-12}$ and $2.7\times 10^{-13}$ $\mu $m$^{2}$, respectively.  This
subtle difference is presumably due to the magnetic fluctuation of the fixed
layer.  Remarkably, the peak positions are roughly symmetric with respect to
$H_{c}$ (namely, peaks A and B, and peaks C and D).  Finally, these symmetric
behaviors with respect to $H_{c}$ are always observed on all three samples.  We
will discuss the implications of these observations later in Sec.~IIID again.
\begin{figure}[tp]%%----------------------------------
\center \includegraphics[width=.70\linewidth]{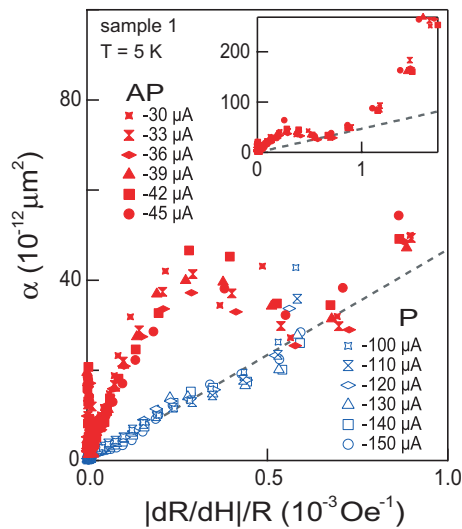} \caption{ $\alpha$ at
several bias currents vs $|dR/dH|/R$ for P (open symbols) and AP (solid
symbols). The inset shows the full scale of the same data for AP. The dashed
line is a guide to the eye.}
\end{figure}

The influence of the magnetic field sweep rate on the MR curve is shown in
Fig.~4, where $H_{c}$ is also a relevant parameter.  The resistance measured by
each sweep rate is plotted as the difference of the data from that obtained with
the slowest sweep rate (50 Oe/min), where the field was ramped from $-$500 to
240 Oe for AP and from 500 to $-$300 Oe for P configurations.  Although
initially there is no sweep rate dependence on the resistance for either
configuration, a sweep rate dependence appears as the magnetic field crosses
$H_{c}$.  This observation strongly indicates that the free electrode feels zero
effective field at $H_{c}$.  The shift of $H_{c}$ from zero field is explained
by taking a magnetic interaction between the free and fixed layers into account.

\subsection{$1/f$ noise}
Now, we discuss what we can learn from the $1/f$ noise.  Previously, Ingvarsson
\textit{et al}.~\cite{AlOnoise3} scaled $\alpha $ versus $|dR/dH|/R$ and
explained the result in terms of a fluctuation-dissipation (FD) relation.  The
scaling of the $1/f$ noise by the FD relation has been tested against
Al$_{2}$O$_{3}$-based~\cite{AlOnoise3,AlOnoise4} and MgO-based~\cite{MgOnoise9,MagLoss}
MTJs with micron-sized junction areas.  Here, we test the scaling by the FD
relation on the free layer of MgO-based MTJs with submicron-sized junctions
following Ingvarsson \textit{et al}.  We replot the data in Figs.~3(c) and 3(d)
and obtain the relation between $\alpha $ and $|dR/dH|/R$ as shown in Fig.~5.
Remarkably, the values of $\alpha $ for the P configuration (open symbols) are
almost proportional to $|dR/dH|/R$.  For the AP configuration, the $\alpha $
values (solid symbols) first deviate from the dashed line (drawn to guide the
eye) because of peaks A and C, and then fall onto the dashed line again.  In the
inset of Fig.~5, $\alpha $ for the AP configuration is shown over the wider
range of $|dR/dH|/R$.  After dropping onto the dashed line, $\alpha $ for AP
rapidly increases with $|dR/dH|/R$ beyond $\sim $1~mOe$^{-1}$.  In this region, the
measured spectra exhibit a complicated enhancement of Lorentzian character
possibly from several RTNs.

If one assumes thermal equilibrium, then the FD relation for this magnetic system is
given by~\cite{AlOnoise3}
\begin{eqnarray}
S_{m}=\frac{2k_{B}T}{\pi \mu _{0}f} \chi _{m}^{\prime\prime}\left(f\right)
\label{eq:five},
\end{eqnarray}
where $S_{m}$, $\mu _{0}$, and $\chi _{m}^{\prime\prime}$ denote, respectively,
the spectral density of magnetic fluctuation, the vacuum permeability, and the
imaginary part of the magnetic susceptibility.  Then by using a typical equation
of $1/f$ noise [Eq.~(3)] and the Kramers-Kronig relation, Eq. (5) reduces
to~\cite{AlOnoise3}
\begin{eqnarray}
\alpha ^{\ast }=\frac{k_{B}T\textrm{A}\Delta R}{2m\mu _{0}R\ln \left(f_{max}/f_{min}\right)}\left(\frac{1}{R}\frac{dR}{dH}\right)
\label{eq:eight},
\end{eqnarray}
where 2$m$ and $\Delta R$ are the respective changes of the magnetic moment and
resistance associated with the reversal of the free layer.  Thus, the measured
linear relation between $\alpha $ and $|dR/dH|/R$ for the P state in Fig.~5 can
be explained by Eq. (6); namely, the origin of $1/f$ noise is thermal agitation
of the magnetic moment of the free electrode. The dashed line corresponds to
Eq. (6), where we take the saturation magnetization $M_{S}$ of CoFeB to be
$1.3\times 10^{6}$ A/m.~\cite{MgOnoise9} From rough estimation, we obtained
$f_{max}/f_{min}\sim 10^{10\pm 2}$, which is consistent with the result
$f_{max}/f_{min}\sim 10^{9}$ that Ingvarsson \textit{et al.}~\cite{AlOnoise3}
reported.
The nonlinear behavior with $|dR/dH|/R$ beyond $\sim $1~mOe$^{-1}$ is
possibly caused by deviation from the thermal equilibrium state of the free
layer.  In fact, the measured dc resistance corresponding to this region
strongly depends on the field sweep rate (see Fig.~4).
According to previous works,~\cite{MgOnoise9,MagLoss2,MagLoss3} the linear relation between $\alpha $ and $|dR/dH|/R$ corresponds to constant magnetic losses and has been shown by Stearrett \textit{et al.}~\cite{MagLoss} in the P and AP configurations.

\subsection{Random telegraph noise }
\begin{figure}[tp]%%----------------------------------
\center \includegraphics[width=.90\linewidth]{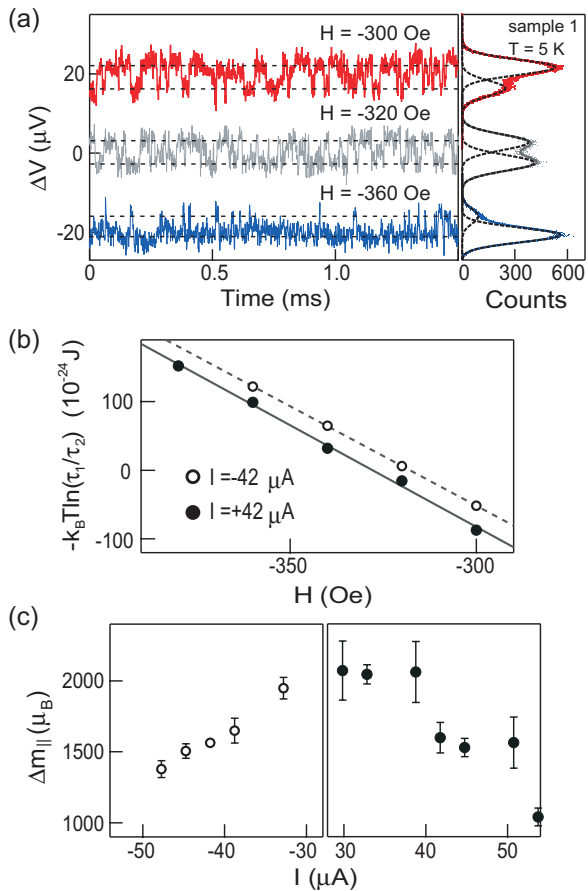} \caption{(a) Magnetic
field dependence of the voltage fluctuations (offset for clarity) measured
around peak A in Fig. 3(b) (left panel) and the corresponding histograms (dots)
($2\times 10^{5}$ points) (right panel). The solid and dashed curves are the
fitted curves (double Gauss function) and each Gaussian component. (b) Logarithm
of the rate of the dwell times for $I=-42$~$\mu $A (open circle) and 42 $\mu $A
(solid circle) as a function of $H$. The dashed and solid lines are the results
of the linear fitting. (c) Bias dependence of estimated fluctuation magnetic
moments.}
\end{figure}
Previously magnetic RTN have been observed in
Al$_{2}$O$_{3}$-based~\cite{AlOnoise3,AlOnoise4} and
MgO-based~\cite{MgOnoise4,MgOnoise6,MgOnoise10} MTJs, which are typically
sensitive to the magnetic field and bias voltage.  Although it is difficult to
deal with the RTN systematically in general, a few authors successfully
estimated the effective magnetic moment of the fluctuator and discussed possible
origins of magnetic RTN.  In Al$_{2}$O$_{3}$-based MTJs with micron-sized
junctions, Ingvarsson \textit{et al.}~\cite{AlOnoise3} and Jiang \textit{et
al.}~\cite{AlOnoise4} suggested a small rotation of a single domain or domain
wall hopping between pinning sites.  Recently, Herranz \textit{et
al.}~\cite{MgOnoise10} reported that the magnetic RTN is caused by magnetic
inhomogeneities and domain walls in free and fixed layers.

We observed a strong enhancement of the Lorentzian components in regions with
specific bias voltages and magnetic fields.  Here we consider peaks A, B, C, and
D in Fig. 3(b) as typical examples.  Generally, the Lorentzian component in the
noise spectral density indicates that the noise is caused by the fluctuation
between two levels, namely, RTN.  The real-time voltage signals near peak A for
$H=-300$, $-$320, and $-$360 Oe are shown in the left panel of Fig.~6(a); the
right panel shows the corresponding histograms for the voltage signals ($2\times
10^{5}$ points for 200 ms).  These signals have two distinct voltage levels, and
hence we attribute the measured Lorentzian components to RTN.  The histograms
are fitted by a double-Gaussian function, which is shown in the same panel by
the dashed lines.  Let us call the two states ``1" and ``2."  From the area of
each Gaussian, which is proportional to the dwell time of each state ($\tau
_{1}$ and $\tau _{2}$), we estimate the ratio of the dwell times between these
states ($\tau _{1}/\tau _{2}$).  Figure 6(b) shows a logarithmic plot of $\tau
_{1}/\tau _{2}$ versus $H$ for $I=\pm 42 $~$\mu $A.  Remarkably, $\log \tau
_{1}/\tau _{2}$ is proportional to the magnetic field.  The strong dependence of
the dwell time for each state as a function of magnetic field indicates that the
RTN is due to a magnetic fluctuator.

We assume that two states with dwell times of $\tau _{1}$ and $\tau _{2}$ have
activation energies of $E\pm \Delta \textbf{m}\cdot \textbf{H}$.  By further
assuming the Arrhenius relation, we express the dwell times $\tau _{1}$ and
$\tau _{2}$ as
\begin{eqnarray}
\frac{1}{\tau _{1,2}\left(H\right)}=\frac{1}{\tau _{0}}\exp \left(-\frac{E\pm \Delta \textbf{m}\cdot \textbf{H}}{k_{B}T}\right)
\label{eq:nine},
\end{eqnarray}
where $E$, $\Delta \textbf{m}$, and $1/\tau _{0}$ are the field-independent
activation energy, the total magnetic moment of the fluctuator, and an attempt
frequency, respectively (where we take + for the 1 state and $-$ for the 2
state).  To estimate the effective magnetic moment parallel to $H$ ($\Delta
m_{\parallel }$), we take the ratio of each dwell
time:~\cite{AlOnoise3,AlOnoise4,MgOnoise10} $\tau _{1}/\tau _{2}\propto \exp
\left(-2\Delta m_{\parallel }H/k_{B}T\right)$.  By a linear fit of $\ln
\left(\tau _{1}/\tau _{2}\right)$ versus $H$ [see Fig. 6(b)], we estimated the
effective magnetic moment $\Delta m_{\parallel }$ to be $1.6\times 10^{3}\mu
_{B}$ for both positive and negative currents ($I=\pm 42 $~$\mu $A), where $\mu
_{B}$ is the Bohr magneton.  This value corresponds to 0.08\% of the total
magnetic moment of the free electrode ($2.0\times 10^{6}\mu _{B}$).  However,
the change in resistance $\Delta R$ resulting from RTN was estimated from the
real-time voltage signal to be 0.13 $\Omega $, which is about 0.035\% of the
total resistance change for the reversal of the free layer, which is consistent
with the above value (0.08\%).  This observation strongly suggests that the
observed RTN is responsible for the magnetic fluctuator.  The bias dependence of
the estimated $\Delta m_{\parallel }$ is shown in Fig.~6(c).  With increasing
bias current, $\Delta m_{\parallel }$ is monotonically decreased for both bias
polarities.  This means that as the injection current gets higher, smaller
magnetic fluctuations are allowed.
\begin{figure}[tp]%%----------------------------------
\center \includegraphics[width=.80\linewidth]{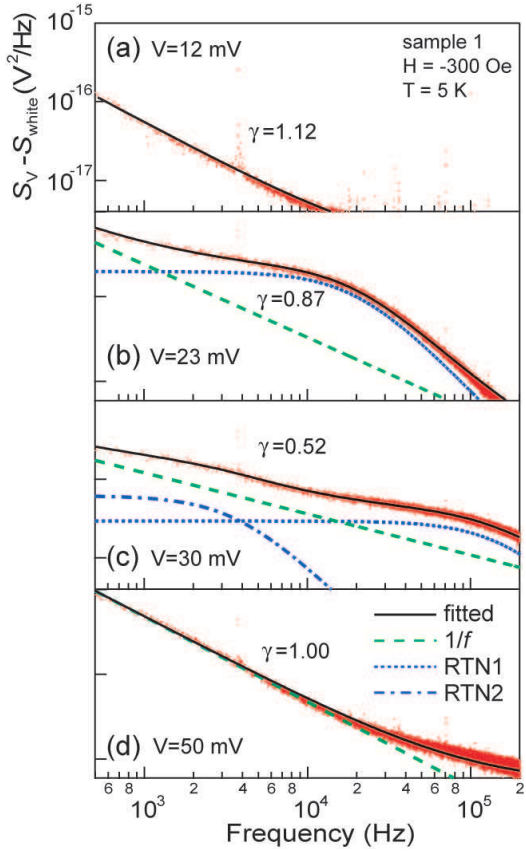} \caption{(a)--(d) The
frequency-dependent component of the spectrum for $V=12$, 23, 30, and 50 mV,
respectively. The solid curves are the fitted curves obtained by summing the
$1/f$ noise, RTN1, and RTN2, where each component is shown as a dashed
curve.}
\end{figure}

The $\Delta m_{\parallel }$ values in the previous
reports~\cite{AlOnoise3,AlOnoise4,MgOnoise10} are larger than those in our
result by at least two orders of magnitude.  Here we discuss the origin of the
fluctuator that contributes to the measured magnetic RTN.  If one assumes full
reversal of a single domain in the free layer, a typical area size of the
magnetic fluctuator is $5.9\times 10^{-6}\;\mu \text{m}^{2}$, where the junction
area is $7.4\times 10^{-3}\;\mu \text{m}^{2}$.  This small value and the absence of
any step in the magnetic hysteresis loops in Fig.~4 indicate that the
contribution of a macroscopic domain wall can be ruled out.  The estimated
$\Delta m_{\parallel }$ and $\Delta R$ for the AP configuration have almost the
same percentages against the full reversal of the free layer.  Finally,
enhancement of magnetic RTN is observed for both configurations.  Based on these
results, we attribute the fluctuator to two quasistable single-domain states
with some strain in the free layer.~\cite{SekiAPL2011}

\subsection{Crossover from the $1/f$ noise to RTN}
\begin{figure}[tp]%%----------------------------------
\center \includegraphics[width=.85\linewidth]{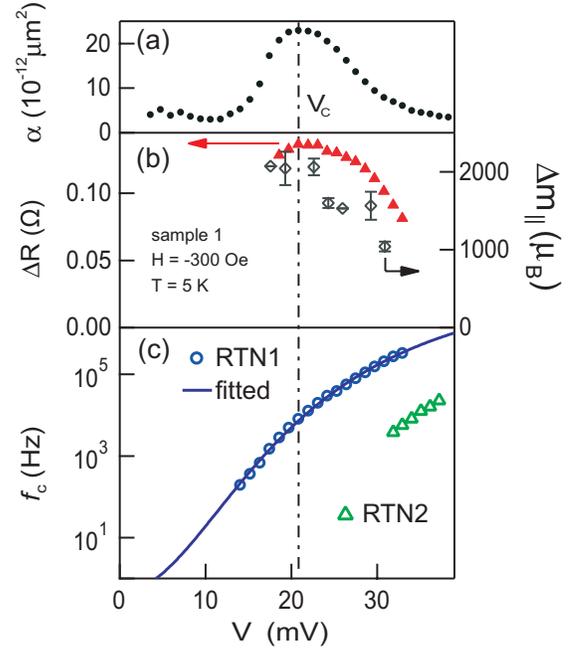} \caption{ Bias
dependence of the parameters characterizing the RTN at $-$300 Oe for AP. (a)
Bias voltage dependence of $\alpha $. (b) The triangle is the resistance change
of RTN1. The open diamond is the estimated fluctuating magnetic moments in
Fig. 6(c). (c) $f_{c}$ for RTN1 (circle) and RTN2 (triangle). The solid curve is
the fit to $f_{c}$ for RTN1 }
\end{figure}
To understand the excitation mechanism of the RTN we focus on the bias
dependence of the noise property at peak A.  Figure 7 represents the
frequency-dependent component of measured spectra at $-$300 Oe (AP
configuration) for $V=12$, 23, 30, and 50 mV.  Although pure $1/f$ noise is
observed in the low-bias region as shown in Fig.~7(a), RTN is dominant at $V=23$
mV, resulting in the strong enhancement of the Lorentzian component (``RTN1")
[Fig. 7(b)].  The characteristic frequency of the RTN is given by the full width
at half maximum of the Lorentzian ($f_{c}$).  For example, $f_{c}=20$~kHz for
RTN1 in Fig.~7(b).  With increasing bias voltage, the characteristic frequency
of RTN1 increases and another Lorentzian component (``RTN2") with different
$f_{c}$ appears [Fig. 7(c)].  Finally, the $1/f$ noise becomes dominant again in
the spectrum [Fig. 7(d)].  To fit the obtained spectra as a summation of $1/f$
noise ($S_{\rm{1/f}}$) and a few Lorentzians ($S_{\rm{RTNi}}$), we use $\beta
/f^{\gamma }$ and $\beta_{i}^{\ast }/[1+(f/f_{ci})^{2}]$, respectively.  Here
$\gamma$ and $f_{ci}$ are the spectral exponent and characteristic frequency of
the Lorentzian for ``RTN$i$" ($i=1$ and 2).  Outside the peak region, the
estimated $\gamma $ is close to 1, whereas it is reduced from 1 as the
Lorentzian components are enhanced.  By taking the Dutta-Dimon-Horn
model,~\cite{Dutta} in which the $1/f$ noise is the result of a superposition of
many RTNs with a broad distribution of activation energies, this suppression of
$\gamma $ accompanied by the enhancement of RTN can be explained by a change of
the energy distribution of the magnetic fluctuators.

The bias dependence of the parameters characterizing RTN at $-300$ Oe for the AP
configuration is summarized in Fig. 8.  $V_{c}\sim 21$~mV is the bias voltage
where $\alpha $ shows its maximum.  The effective magnetic moment of the
fluctuator, $\Delta m_{\parallel }$, and the change in resistance, $\Delta R$,
are compared in Fig. 8(b).  The behaviors of $\alpha $, $\Delta m_{\parallel }$,
and $\Delta R$ are similar to each other in that they have their maxima at $V_{c}$
and then decrease as the bias is increased.  Finally, both $\Delta m_{\parallel
}$ and $\Delta R$ become almost half of their maxima at the bias at which the
peak of $\alpha $ disappears.  Figure 8(c) shows that, although the estimated
$f_{c}$ for RTN1 and RTN2 exhibits almost an exponential increase with the
bias for their whole curves, there is a kneelike structure in the curve for
RTN1 at $V_{c}$.  We fit $f_{c}$ for RTN1 to $f_{c}=f_{0}\exp
\left(-E/k_{B}T^{\ast }\right)$ using the Arrhenius relation and taking Joule
heating into account, where $f_{0}$ is an attempt frequency.  We also assume
that the activation energy $E$ is independent of the bias voltage and that the
effective temperature $T^{\ast }=T+\zeta V^{2}$.  The solid curve in Fig. 8(c)
corresponds to the fitted curve, where the estimated $E$ and $\zeta $ are
$1.3\times 10^{-21}\text{ J}$ and $1.3\times 10^{4}\text{ K}/\text{V}^{2}$,
respectively.  Comparing this result to the previous one,~\cite{AlOnoise6} in
which the activation energy for nonmagnetic RTN in Al$_{2}$O$_{3}$-based MTJs
was found to be $0.3\times 10^{-19}\text{ J}$, we see that our value is much
smaller.  This is consistent with the absence of nonmagnetic RTN in high-quality
MTJs in the low-temperature, low-bias regime.

Returning to the magnetic configuration dependence of the RTN, one sees that
$\alpha $ values for peaks A and B are enhanced at roughly symmetric positions
with respect to $H_{c}$ [see lower panels in Figs. 3(b) and 3(c)].  The values
of $\alpha $ for peaks A and B have their maxima at almost the same bias
voltage, whereas the amplitude of the maximum for peak A is almost seven times
as large as that of peak B.  We assume that the same magnetic fluctuator
contributes to these two peaks.  The symmetric hysteresis loop of the free layer
with respect to $H_{c}$ also supports this assumption (see Fig.~4).  Here, the
difference in the peak amplitude between A and B cannot be accounted for by
conventional Joule heating.  That is, because the Joule heating effect is
proportional to the bias current at a fixed bias voltage, this effect for peak A
is only 1/3 that for peak B.  These results may imply a spin-dependent heating
process of a localized spin system in the free layer depending on the magnetic
configurations.

\begin{figure}[tp]%%----------------------------------
\center \includegraphics[width=.80\linewidth]{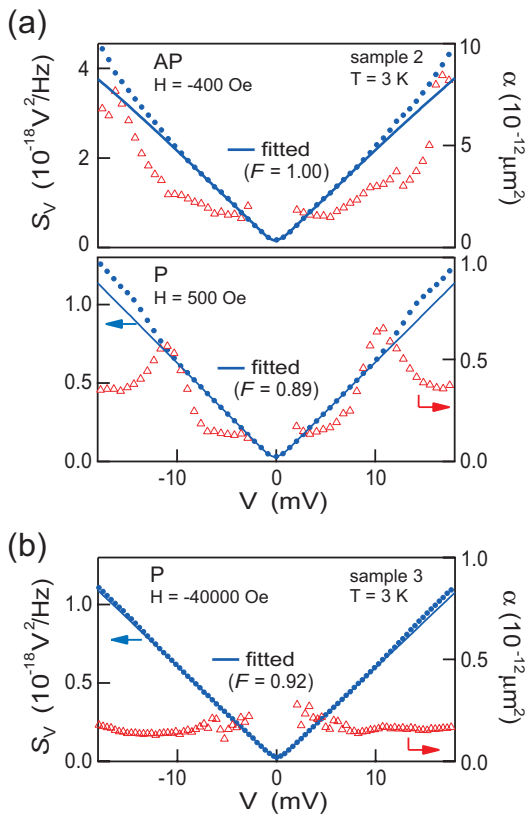} \caption{(a) White noise
component and $\alpha $ measured at 3~K in the P ($H=500$ Oe) and AP ($H=-400$ Oe) configurations on
sample 2 as a function of $V$. The solid curve is the fitted curve between
$V=-10$ and 10 mV. (b) Counterpart of Fig. 9(a) in the P configuration
($H=-40,000$ Oe) on sample 3. }
\end{figure}
\subsection{Shot noise}
Finally, we discuss the shot noise to connect our previous work with the present
one and to support further evidence of our claim made before.  The
frequency-independent component of the spectrum is well described by Eq.~(1).
$F$ describes how the noise deviates from the Poissonian value and thus
characterizes the partition process of the electron tunneling.  Conventional
tunnel junctions exhibit $F=1$, reflecting the Poissonian process.  Regarding
Al$_{2}$O$_{3}$-based MTJs, after the first report of the full shot noise
($F\sim 1$),~\cite{AlOnoise4} reduced Fano factors ranging from 0.45 to 1 were
reported.~\cite{MTJshot1,MTJshot2} This reduction can be explained by the
sequential tunneling model,~\cite{MTJshot2} where the process of two-step
tunneling through impurities within the barrier is assumed.  In this theoretical
model, $F$ strongly depends on the asymmetry of the each tunneling and can be
0.5 to 1.

Full shot noise in MgO-based MTJs was reported by Guerrero \textit{et al}. and
some of the authors of the present paper in experiments where the MgO barriers
were as thick as 3 and 1.5 nm, respectively.~\cite{MTJshot3,MTJshot4} This
indicates that MgO-based MTJs are free from the process through impurity sites,
possibly resulting from the high quality of the crystallized MgO.  Later, we
reported the suppression of the Fano factor for the P configuration (typically
0.91) with a 1.05-nm-thick MgO barrier, whereas the $F$ for the AP configuration
is almost unitary (typically 0.99).~\cite{MTJshot6} To explain this subtle
reduction of $F$ within the above sequential tunneling model, we have to assume
a very asymmetric barrier (1:100), which is unrealistic as the barrier of our
MTJ is thin.  Moreover, by using this model, we cannot explain the magnetic
configuration dependence of the reduced $F$, and hence we can rule out this
scenario.  We note that all of the experimental results for $F$ values including
the thick-barrier case can be explained by recent theoretical work based on a
first-principles calculation.~\cite{MTJshot7} Namely, the Fano factor, which is
reduced from 1, is a direct consequence of the Pauli exclusion principle,
signaling that there are coherent channels with high transmission probabilities
through the epitaxial MgO barriers.  Thus, our result on the shot noise gives
unique evidence for coherent tunneling through a crystallized MgO barrier.

We show in Fig.~9 the typical results of the white-noise component $S_{V} $ and $\alpha $ at 3 K as a function of bias voltage.
In Fig.~9(a), $S_{V} $ is nicely fitted by Eq.~(2) for $ |V| \le 10$ mV.
For $| V| > 10$ mV, $\alpha $ increases due to enhancement of RTN for both configurations, and, as a result, $S_{V} $ deviates from the fitted curve.
%For $| V| > 10$ mV, $\alpha $ has peaks caused by RTN for both bias directions; therefore $S_{V} $ deviates from the fitted curve outside the peaks.
It is noted that the effect of the $1/f$ noise on $S_{V} $ is negligibly small in this bias range.
The above observation indicates that to accurately estimate the Fano factor immune to the frequency-dependent noise, it is necessary to analyze the noise in the low-bias regime at low temperature.
Experimentally, we can suppress magnetic RTN by stabilizing the magnetization by applying a large magnetic field for the P configuration.
In fact, in Fig.~9(b), the enhancement of RTN is suppressed, and thus $S_{V} $ is almost perfectly fitted by the shot-noise formula with $F=0.92$.

\section{Conclusion}
The low-frequency noise properties in submicron-sized CoFeB/MgO/CoFeB-based MTJs
with thin tunneling barriers are systematically investigated.  The measurements
are carried out at low temperature by controlling the magnetic field and bias
current (voltage), where we focus on the noise properties on the magnetic
hysteresis loops of the free layer.  A clear correlation between $\alpha $ and
$|dR/dH|/R$ is observed.  The nice scaling of the observation by the FD relation
indicates that the main origin for the $1/f$ noise is thermal magnetic
fluctuation of the free layer.  RTN is observed inside the magnetic hysteresis
loops for both configurations.  We found that this is due to the magnetic
fluctuation between two quasistable single-domain states with some strain in the
free layer.  Although the $1/f$ noise is almost independent of the magnetic
configuration, RTN is remarkably enhanced for the AP configuration and at
specific bias voltages.  Such results indicate a spin-dependent activation
process of RTN.  Shot noise measurement gives us quantitative information for
coherent tunneling.

Our study shows that a systematic study on the noise of MTJs is possible by
using devices with well-crystallized thin MgO barriers and by measuring the
noise at low temperature and low bias.  Further study of the noise properties in
MgO-based MTJs with various barrier thicknesses is necessary to systematically
elucidate the mechanism of the noise and improve the device properties.

\section*{Acknowledgment}
This work was partially supported by the JSPS Funding Program for Next
Generation World-Leading Researchers.

%\section{ref}

\end{document}